# Confined-state physics and signs of fermionization of moiré excitons in WSe$_2$/MoSe$_2$ heterobilayers

F. Lohof[1*], J. Michl[2*], A. Steinhoff[1], B. Han[3], M. von Helversen[5], S. Tongay[6],
K. Watanabe[7], T. Taniguchi[8], S. Höfling[2], S. Reitzenstein[5], C. Anton-Solanas[9],
C. Gies[1], and C. Schneider[3,4]

[1] Institute for Theoretical Physics and Bremen Center for Computational Material Science, University of Bremen, 28359 Bremen, Germany
[2] Julius-Maximilians-Universität Würzburg, Physikalisches Institut and Würzburg-Dresden Cluster of Excellence ct.qmat, Lehrstuhl für Technische Physik, Am Hubland, 97074 Würzburg, Deutschland
[3] Institute of Physics, University of Oldenburg, 26129 Oldenburg, Germany
[4] Center of Nanoscale Dynamics, University of Oldenburg, 26129 Oldenburg, Germany
[5] Institut für Festkörperphysik, Technische Universität Berlin, Hardenbergstrasse 36, 10623 Berlin, Germany
[6] School for Engineering of Matter, Transport, and Energy, Arizona State University, Tempe, Arizona 85287, USA
[7] Research Center for Functional Materials, National Institute for Materials Science, 1-1 Namiki, Tsukuba 305-0044, Japan
[8] International Center for Materials Nanoarchitectonics, National Institute for Materials Science, 1-1 Namiki, Tsukuba 305-0044, Japan
[9] Depto. de Física de Materiales, Instituto Nicolás Cabrera, Instituto de Física de la Materia Condensada, Universidad Autónoma de Madrid, 28049 Madrid, Spain

[*]Both authors have contributed equally to this work.

E-mail: flohof@itp.uni-bremen.de

**Abstract**

We revisit and extend the standard bosonic interpretation of interlayer excitons in the moiré potential of twisted heterostructures of transition-metal dichalcogenides. In our experiments, we probe a high quality MoSe$_2$/WSe$_2$ van der Waals bilayer heterostructure via density-dependent photoluminescence spectroscopy and reveal strongly developed, unconventional spectral shifts of the emergent moiré exciton resonances. The observation of saturating blueshifts of successive exciton resonances allow us to explain their physics in terms of a model utilizing fermionic saturable absorbers. This approach is strongly inspired by established quantum-dot models, which underlines the close analogy of interlayer excitons trapped in pockets of the moiré potential, and quantum emitters with discrete eigenstates.

## 1. Introduction

Semiconductor van der Waals heterostructures are a recognized platform for emerging applications in nanophotonics and optoelectronics. Band alignments of different material combinations allow for tunability of the emission wavelength of spatially-indirect excitons that are made up from electrons and holes residing in the individual layers of the bilayer system. The possibility to influence the strength of the Coulomb interaction by dielectric [1]–[4] and strain [5]–[7] engineering has created additional tuning knobs that can be em-ployed to design optoelectronic building blocks with tailored absorption and emission properties [8]–[10].

Controlling the twist angle between monolayers of transition metal dichalcogenides (TMD), forming homo- or heterobilayer systems has opened up yet another avenue of solid-state emitter nanoengineering [11]–[14]. At small twist angles between the stacked TMD monolayers, an additional potential landscape is superimposed that, in the ideal case, forms a periodic lattice of confinement potentials, both for interlayer excitons (ILX) as well as intralayer excitons [15]–[17]. ILX traversing the moiré potential are referred to as indirect moiré excitons – seemingly bosonic entities that are made up of a charge carrier in each layer that strongly experience the moiré potential [18]–[20] with the addition of featuring a permanent dipole moment [21]–[24]. Moiré excitons have been observed in a series of papers, which reported the emergence of multi-peak features associated with excitonic bands [25], as well as quantum dot-like signatures associated with full excitonic trapping yielding a quantum blockade [16], [26].

Here, we revisit the core nature of moiré excitons at small twist angles and provide indications that their non-linearity is centrally governed by fermionic saturation, which is very much analogous to discrete-state quantum-dot arrays [20], [27]–[30]. By describing results of pump-power dependent spectroscopic measurements with a configuration-based model, we explain the appearances and line shifts of two moiré ILX resonances in terms of phase-space filling of saturable absorbers.

## 2. Sample & spectroscopy experiments

Fig. 1(a) depicts an optical microscope image of the studied sample. It is composed of an AlAs/AlGaAs distributed Bragg reflector with 20 $\lambda/4$ mirror pairs, which is capped by a thin layer of SiN to support the electro-magnetic field maximum at the SiN-air interface. We utilize the dry stamping method to assemble the van-der-Waals heterostructure composed of a thin hBN layer (approx. 5 to 10 nm), a monolayer of $WSe_2$, a $MoSe_2$ monolayer and finally a second hBN layer (approx. 5-10 nm), which were prepared by mechanical exfoliation. Finally, the sample stack with a moiré twist angle of 59° was annealed for 3h at 300°C in inert gas atmosphere.

We study the basic optical properties of our device via position-resolved scanning micro-photoluminescence (microPL) spectroscopy at a cryogenic temperature of 5 K. The sample was excited by a frequency doubled Nd.YAG laser (532 nm).

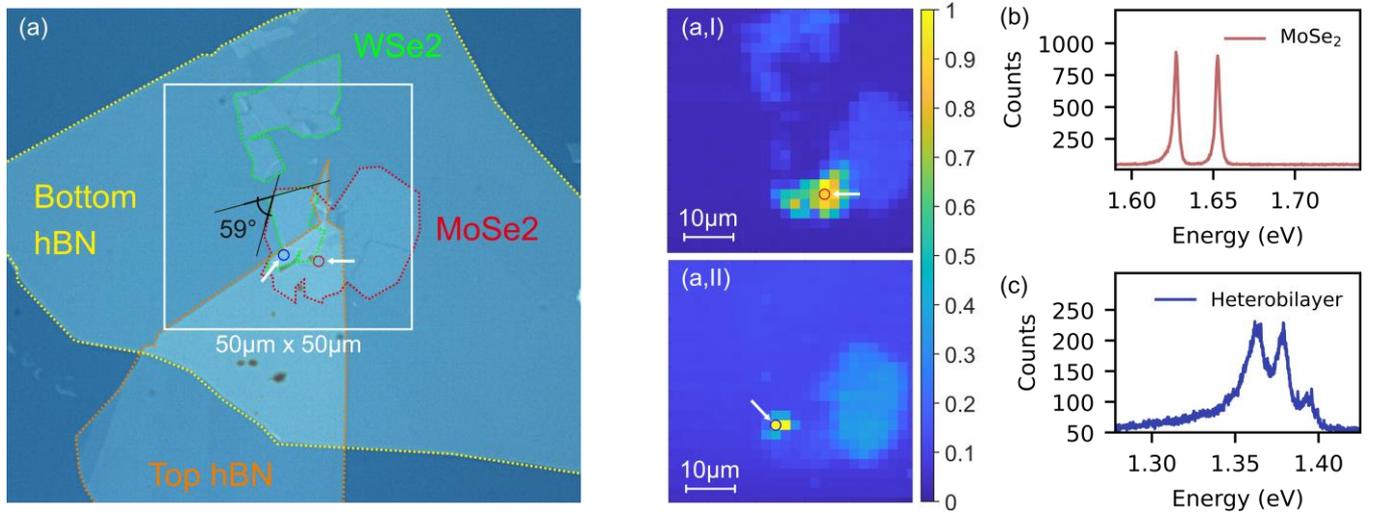

*Figure 1. (a) Microscope image of the sample stacking with a bottom and top layer of hBN (limits highlighted in dashed yellow and orange, respectively, as a guide to the eyes) and the two stacked monolayers of $WSe_2$ (green) and $MoSe_2$ (red). The region of interest is indicated by a white square of 50 μm side. The twist angle between the monolayers is ~59°. (a,I/II) Real-space PL map (in linear scaling) in the region of interest integrating the energy range between 1.647-1.662/1.312-1.401 eV, respectively. In panel **a,I/a,II**, the specific spot encircled in red/blue color is highlighted in the heterobilayer area (and pointed at with a white arrow), where the strongest emission is detected at the respective wavelength range. (b,c) Corresponding spectrum extracted in the chosen pixel, depicted accordingly in the same red/blue color. The excitation conditions are P= 0.5 μW excitation power and 5s integration time.*

In the spectral region between 1.6–1.68 eV, we clearly observe the characteristic emission features of the MoSe$_2$ monolayer (cf. Fig. 1(a,I)). It is worth noting that the monolayer is fully encapsulated in the bottom region of the sample plotted in Fig. 1(a), whereas it is uncapped in the spatial region on the right-hand side. The optical spectrum in the encapsulated region features two distinct PL peaks (cf. Fig. 1(b)) which we attribute to the PL from the exciton- and trion resonances, and which feature spectral linewidths as narrow as 2.5 meV with a Lorentzian lineshape. This clearly reflects the very high optical quality of our device. In Fig. 1(c), we depict the PL spectrum in the spectral region 1.3–1.41 eV. This spectral signature evolves in the region of the full heterobilayer and is dominated by two strong PL features at 1.362 eV and 1.378 eV, with a much dimmer feature at 1.4 eV. We note that although the spectra in Fig. 1(b),(c) are recorded close the edge of material regions, we excluded the influence of strain at the monolayer edge as the spectral features are consistent in terms of energy position and linewidth over the whole mapped-out area.

The energy separation of peak 1 and 2 is 18 meV, which is consistent with previously reported assignment of singlet and triplet ILX trapped in $H^h_h$ atomic registry of high-quality 2H stacked WSe$_2$/MoSe$_2$ heterobilayer [31]. The total energy shift of ~50 meV of both peaks 1 and 2, relative to the current literature, could be due to a reduced moiré potential and local strain in our sample, in which the former one is twist-angle dependent. In Ref. [32] the peaks are energetically consistent with our results, and they are assigned to the $H^h_h$-site singlet ILXs with finite momentum. More generally, these emission peaks can be associated with the ILX, which is spatially confined in the high-symmetry $H^h_h$ potential minima of the moiré superlattice in the case of H-stacking here.

Interestingly, as opposed to reference samples that were prepared without the encapsulation process, the overall spectrum is dominated by the PL signal, which we associate with the indirect excitonic spectral features, whereas the single-layer PL features are strongly quenched, reflecting an efficient charge transfer into the low-lying ILX state at significantly shorter timescales than the recombination from the direct exciton states [12], [33]–[35].

Thanks to the charge separation in the two monolayers, ILX yield a permanent dipole moment, which has been assessed to approximately 0.5 nm e (e is the elementary charge) [8]. As a consequence, as the exciton density is modified externally, the spectral features are expected to experience a repulsive shift due to excitonic dipolar interaction [21]–[24]. This effect can be studied using power-dependent measurements, which we have performed and show in Fig. 2(a). With a laser spot diameter of 2 µm, the driving-power density is varied by four orders of magnitude from $3.1 \cdot 10^{-3}$ µW/µm$^2$ to $3.1$ µW/µm$^2$. In this range of excitation powers, we observe a strong blueshift of all excitonic reso-nances by 10 to 20 meV. This behavior becomes even more distinct in the representation in Fig. 2(b), where the vertical axis is scaled logarithmically, and the peaks are normalized to unity. This closer inspection reveals that the individual spectral lineshifts behave differently: The emission feature at an energy of 1.353 eV, named ILX1, displays the most profound energy shift at lowest pump powers and saturates at a pump power of ~ 1 µW. In turn, the second peak, emergent at 1.37 eV and labelled ILX2, only displays a modest energy shift for pump power below 1 µW. At higher pump powers, the slope of the blueshift of ILX2 becomes larger than that of ILX1. A detailed quantitative discussion of this opposing behavior is given in Section 4 in the context of Fig. 4. In Fig. 2(c), we plot the energy difference between these two features, which develops a clear minimum at a pump power between 1-3 µW, revealing an anti-crossing behavior of the peaks. While such an anti-crossing is canonical for strongly coupled modes (either electronic, magnonic, photonic, or mechanical ones), we claim that in our system, its origin roots in the unique interplay of fermionic saturation and bosonic interactions.

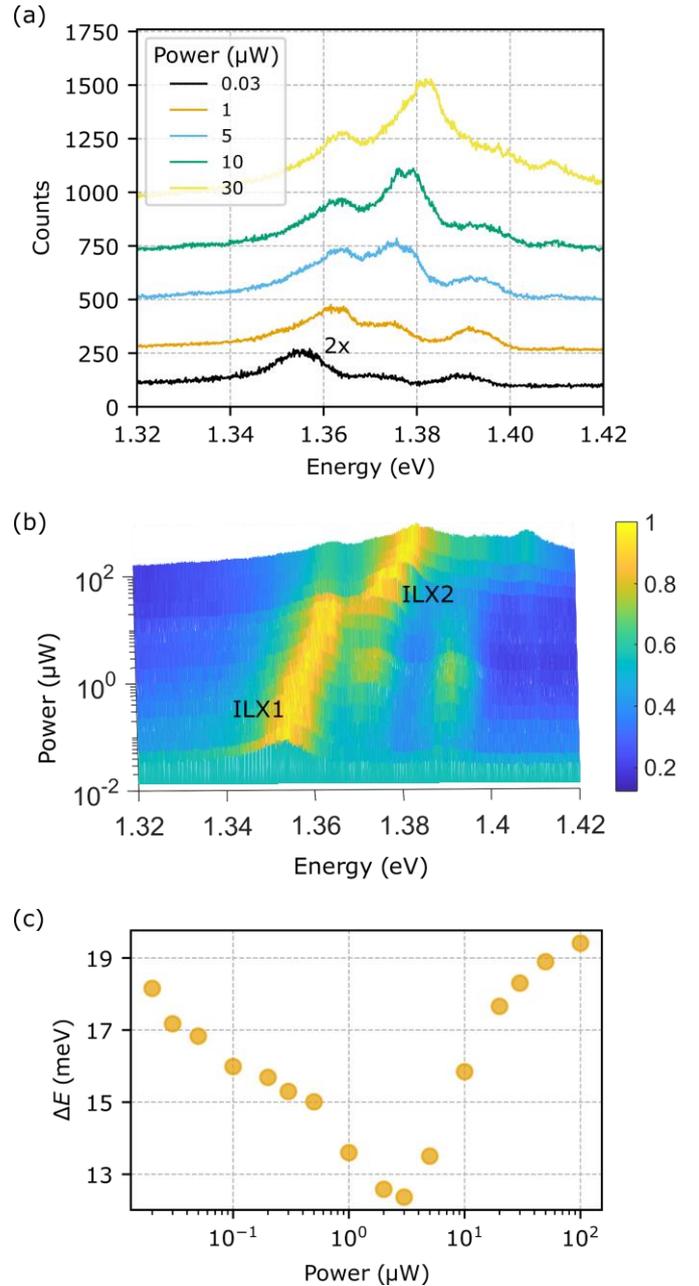

***Figure 2.** (a) PL spectra of the ILX feature, recorded using ascending pump powers. (b) Representation of the same measurement in false color scale. The y-axis represents the excitation power in logarithmic scale. The intensities are normalized to unity. (c) Energy difference between the two main peaks versus excitation power. A minimum energy difference is detected at an excitation power of P= 3 µW. Although reminiscent of an anti-crossing point, we explain the behavior by a successive onset of fermionic saturation of the blueshift.*

## 3. QD model for moiré trapped ILX

To provide a quantitative understanding of the excitation-power dependency of the line shifts of both ILX resonances in our system, we employ a configuration model for the ground and excited states of the lowest-lying ILX residing within a moiré potential pocket. The model assumes two confined exciton states in each moiré unit cell that we label '*s*' and '*p*' in analogy to semiconductor quantum dots [15], [25], [36]. Unlike quantum-dot states, moiré ILX bands exhibit a certain degree of dispersion. However, recent computational results suggest that for twist angles up to 2°, the two lowest ILX states can be safely approximated as dispersion-free [17, 37]. As an ILX is composed of an electron and hole occupying the confined single-particle states, a finite number of configurations is possible for a given number of discrete confined states— these configurations are shown in Fig. 3. It turns out that it is sufficient if we limit ourselves to including those configurations where electron and hole occupy the same single-particle shell.

It is worthwhile pointing out that configurations of a purely bosonic system would strongly differ from the ones shown in Fig. 3, since an arbitrary number of particles could occupy any of the single-particle levels. This is where the fermionic nature of the ILX constituents explicitly comes into play, as the total number of excitations is limited due to the discrete nature of the single-particle states. Nevertheless, higher lying confined states as well as unconfined states exist that we account for in terms of a bosonic reservoir at higher energies. The lowest reservoir energy $\varepsilon_r$ is to be understood as an effective quantity.

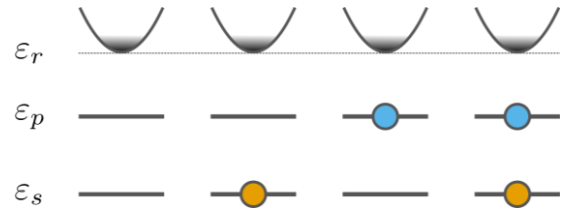

*Figure 3.* Four possible configurations in the quantum-dot model with two bound levels. In the configuration model, we explicitly account for s- and p-like exciton (orange and blue) as well as higher-lying states via an effective bosonic reservoir with parabolic dispersion (grey).

The described microscopic configuration picture is expressed by the following Hamiltonian:

$$H = \begin{pmatrix} 0 & 0 & 0 & 0 \\ 0 & \varepsilon_s & 0 & 0 \\ 0 & 0 & \varepsilon_p & 0 \\ 0 & 0 & 0 & \varepsilon_s + \varepsilon_p \end{pmatrix}. \quad [1]$$

Here, the energies $\varepsilon_s$ and $\varepsilon_p$ correspond to the experimentally observed peaks ILX1 and ILX2 in Fig. 2. In the limit of zero excitation power, they are given by the lowest eigenenergies of excitons confined to a moiré superlattice site. Note that the energies include Coulomb interaction within a moiré site, since the configurations are the exact eigenstates if we assume that configuration interaction can be neglected [38], [39]. The configuration energies are subject to power-dependent renormalization terms that arise from ILX densities in neighbouring sites in the moiré lattice, which are treated in mean-field approximation.

It is well established that the exciton dipolar repulsion gives rise to a blueshift that is approximately linear in exciton density [21]–[24]. Here we account for contributions from interactions between excitons occupying different states,

$$\varepsilon_s = \varepsilon_s^{(0)} + v_{ss}n_s + v_{sp}n_p + v_{sr}n_r, \quad [2]$$
$$\varepsilon_p = \varepsilon_p^{(0)} + v_{sp}n_s + v_{pp}n_p + v_{pr}n_r, \quad [3]$$

where $\varepsilon_s^{(0)}$ and $\varepsilon_p^{(0)}$ are the ILX zero-excitation energies, $n_i$ the exciton occupation numbers, and $v_{ij}$ are effective Coulomb interaction matrix elements responsible for the blueshift. The density matrix describing the configurations in a grand-canonical ensemble picture can be written as

$$\rho = \frac{1}{Z}e^{-\beta(H-\mu N)} \quad \text{with} \quad Z = \text{Tr}[e^{-\beta(H-\mu N)}], \quad [4]$$

where $\beta = 1/(k_B T)$ is the inverse effective ILX temperature and µ the chemical potential. The occupation numbers for ILX in the two quantum-dot levels are found by taking the trace over the single-particle level subspace,

$$n_s = \frac{1}{Z}\left(e^{-\beta(\varepsilon_s-\mu)} + e^{-\beta(\varepsilon_s+\varepsilon_p-2\mu)}\right), \quad [5]$$

$$n_p = \frac{1}{Z}\left(e^{-\beta(\varepsilon_p-\mu)} + e^{-\beta(\varepsilon_s+\varepsilon_p-2\mu)}\right). \quad [6]$$

These occupation numbers reflect fermionic population and are, thus, bounded by a maximum ILX number of 1. Therefore, excess ILX that are excited must be accommodated in higher lying states that are modelled in terms of the bosonic reservoir. Thereby, the number of ILX per moiré lattice site is not bounded, even though only the lowest occupation states are explicitly accounted for in the configuration model. The reservoir's ILX density is given by

$$N_r = \frac{m_r}{\hbar^2 \pi}\int_{\varepsilon_r}^{\infty}\frac{1}{e^{\beta(\varepsilon-\mu)}-1}d\varepsilon = -\frac{m_r}{\hbar^2\pi}\frac{1}{\beta}\ln\left(1-e^{-\beta(\varepsilon-\mu)}\right), \quad [7]$$

where $m_r/\hbar^2\pi$ is the 2D density of states and $m_r$ the effective exciton mass of the reservoir's parabolic dispersion. We define the reservoir's occupation number $n_r = A_M N_r$ as the number of reservoir ILX in the moiré unit cell with area $A_M$. To obtain a numerical solution to this model, the renormalized energies $\varepsilon_s$ and $\varepsilon_p$ are calculated self-consistently as a function of the total density $N_{tot} = (1/A_M)(n_s + n_p + n_r)$, using a relation $N_{tot} = \eta P$ between excitation power $P$ and the effective quantum efficiency $\eta$, and with the effective Coulomb interaction matrix elements $v_{ij}$ as fit parameters. Table 1 provides numerical values for all parameters used in the evaluation of the model.

Our approach includes a treatment of local (on-site) and nonlocal (inter-site) carrier-carrier correlations on different levels of accuracy. An analogy to this procedure is given by dynamical mean-field theory (DMFT) [40]. In DMFT, a many-body lattice problem is mapped to a single-site (impurity) model interacting with a bath of surrounding particles via a frequency-dependent mean field. The mean field is fixed by the self-consistency requirement that the bath describes the same carriers as those on the local site. DMFT can be shown to become more and more accurate with an increasing number of neighbors for each site. We assume that the number of neighboring moiré unit cells is sufficient to justify such a treatment. In our case, a static mean field is applied.

| Temperature | $T$ | 55 K |
|---|---|---|
| ILX1 (s) zero-excitation energy | $\varepsilon_s^{(0)}$ | 1.354 eV |
| ILX2 (p) zero-excitation energy | $\varepsilon_p^{(0)}$ | 1.371 eV |
| Reservoir energy | $\varepsilon_r$ | 1.389 eV |
| Reservoir effective mass | $m_r$ | 0.84 $m_0$ |
| Moiré unit cell area | $A_M$ | 313.4 nm² |
| Effective quantum efficiency | $\eta$ | 6.4 10⁻³ nm⁻²/µW |
| Matrix elements | $v_{ss}$ | 5.48 meV |
| --- | $v_{sp}$ | 2.33 meV |
| --- | $v_{pp}$ | 2.68 meV |
| --- | $v_{sr}$ | 0.011 meV |
| --- | $v_{pr}$ | 0.047 meV |

*Table 1. Effective parameters used in the configuration model. $m_0$ is the free electron mass.*

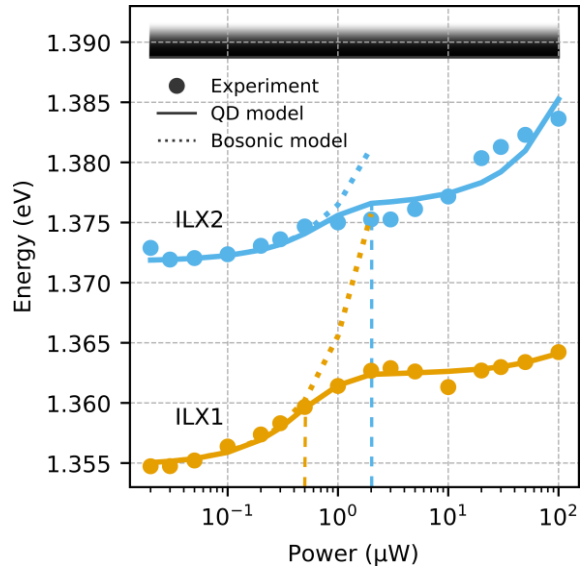

*Figure 4. Peak position of ILX1 and ILX2, extracted from spectra in Fig. 2(b) (circles), and theory predictions from the configuration QD model (solid lines). The ILX reservoir at higher energies is indicated by the shaded area. Vertical dashed lines indicate pump powers where s-state (orange) and p-state (blue) occupations saturate. Dotted lines are calculated with purely bosonic occupation functions and fail to reproduce the observed saturation of blue shifts.*

## 4. Discussion

In Fig. 4, we compare numerical results for the excitation-power dependent energy shifts of the two lowest moiré excitons with the experimental data already shown in Fig. 2. For a wide range of excitation powers, our model reproduces the experimentally observed behaviour very well. At low excitation power up to about 0.2 µW, both resonances undergo an approximately linear blueshift due to the filling of the s-like exciton state (ILX1). The s-exciton shift is significantly stronger than the p-exciton shift (ILX2), which is explained by the dipolar interaction between s-state excitons $v_{ss}$ being stronger than between excitons in different states $v_{sp}$. At elevated excitation powers around 0.5 µW, the s-state occupation starts to saturate (vertical orange line), leading to a corresponding saturation of efficient s-exciton renormalization, while the p-exciton population increases. Since inter-state interaction is relatively weak, this transition from s- to p-state filling leads to a reduced slope of the blueshift of the s-exciton. Around 2 µW excitation power, the p-state saturates as well (blue dashed line), while higher states begin to be populated. The weak interaction of s- and p-states with higher states leads to a significant reduction of blueshifts above 1µW. However, we find that the p-state renormalization in this regime is several times stronger than that of s-states, corresponding to a larger slope of the blueshift. As a comparison we show the results from the same calculation but with purely bosonic occupation functions for the two levels (dotted lines). This model also reproduces the initial linear blueshift, but in the absence of fermionic saturation effects the bosonic occupations of both levels increase indefinitely with growing excitation power, i.e., a purely bosonic exciton picture fails to explain the successive saturation of the blueshift in the two ILX emission features. Hence, the observed non-linear blueshifts of moiré excitons can be fully traced back to discrete state filling effects of saturable absorbers.

The implications of our findings call for a revisited view on moiré excitons: Albeit perfectly bosonic entities at first glance, they are made up from fermionic constituents that cause a deviation from purely bosonic behaviour when confined to a moiré pocket with few discrete localized states. Only recently, the possibility of undergoing a superfluid to Mott transition of moiré excitons in terms of a Bose-Hubbard model has been explored [37], [41], [42]. In this respect, this is where the analogy between ILX in the moiré potential and clouds of ultracold bosonic gases in optical lattices may come to an end.

Furthermore, we believe that it is the fermionic saturation of moiré excitons in discrete levels, as opposed to the dipolar repulsion of trapped bosons, which explains the consistent emission of single photons from moiré trapped excitons [26]. If the ILX were truly behaving more or less like ideal bosonic particles, multiple occupation of the ILX ground state would be more likely to occur even at low temperatures due to the steep slope of the Bose-Einstein distribution function, causing a detrimental effect to any observed antibunching from such emitters.

Finally, we would like to position ourselves to the recent suggestion of a correlation-induced transition between different linear regimes of the dipolar blueshift with increasing excitation density [21]. The underlying picture, which is complementary to the one in this paper, is borrowed from coupled quantum wells [43]. It attributes the weaker blueshift to an onset of classical pair correlations at higher densities, while a homogeneous exciton gas is assumed at low densities. This approach does not take into account the quantum correlations induced by the fermionic substructure of confined ILX. Moreover, for the experimentally observed *simultaneous* lineshifts of both resonances ILX1 and ILX2, we could not obtain a consistent agreement over the considered density range with the model from [43]. This is the main reason why we find phase-space filling effects due to the fermionic contributions of ILXs confined to moiré pockets a more likely explanation.

## 5. Conclusion

With a combination of power-dependent spectroscopic measurements and a configuration-based model that accounts for phase-space filling effects, we provide insight into the lineshift behavior of two moiré ILX resonances with increasing carrier density. Both resonances exhibit linear blueshifts that are offset from one another and transition into saturation, creating an anti-crossing-like signature. Different explanations may be conceivable as explanation, such as an avoided crossing of close-lying bands due to the moiré potential, or correlation-induced deviations from the linear lineshift in the pair-correlation dominated regime of a bosonic exciton gas. Having investigated these options, we conclude that the origin lies in the fermionic nature of the constituents of ILX: Electron and hole forming the ILX occupy the discrete states in the quantum-dot like confinement potential of the moiré pockets, where they are subject to Pauli blocking. In that sense, we interpret our results as successive state-space filling that arises from the fact that ILX, despite their large Coulomb binding energy, are merely composite bosons that may reveal their partly fermionic nature in the right circumstances.

## Data availability statement

The data that support the findings of this study are available upon reasonable request from the authors.

## Acknowledgements

We are grateful for funding by the Deutsche Forschungsgemeinschaft (DFG) via the priority program SPP2244 (projects Gi1121/4-1, Ste2943/1-1, Schn1376/14-1, Re2974/26-1). This project was funded within the QuantERA II Programme that has received funding from the European Union's Horizon 2020 research and innovation programme under Grant Agreement No. 101017733, and with funding from the German ministry of education and research (BMBF) within the project EQUAISE. F. Lohof acknowledges funding from the Central Research Developing Funds (CRDF) of the University of Bremen. B. Han acknowledges the support from Alexander von Humboldt Foundation. C. Schneider gratefully acknowledges funding from the European Research Council (ERC) within the project UnLimIT 2D. Financial support by the Niedersächsisches Ministerium für Wissenschaft und Kultur ("DyNano") is gratefully acknowledged. C. Anton-Solanas acknowledges the support from the Comunidad de Madrid fund "Atraccion de Talento, Mod. 1", Ref. 2020-T1/IND- 19785, and Grant no. PID2020113445-GB-I00 funded by the Ministerio de Ciencia e Innovación (10.13039/501100011033). S. Höfling acknowledges funding by the DFG via the grant Ho5194/16-1. K. Watanabe and T. Taneguchi acknowledge support from JSPS KAKENHI (Grant Numbers 19H05790, 20H00354 and 21H05233). S. Tongay acknowledges primary support from NSF CMMI-1933214, NSF 1904716, NSF 1935994, NSF ECCS 2052527, DMR 2111812, and CMMI 2129412. Support by the DFG (INST 184/220-1 FUGG) is further acknowledged.

## References

[1] M. Rösner, C. Steinke, M. Lorke, C. Gies, F. Jahnke, and T. O. Wehling, "Two-Dimensional Heterojunctions from Nonlocal Manipulations of the Interactions," *Nano Lett.* **16**, 2322, (2016).

[2] M. Florian *et al.*, "The Dielectric Impact of Layer Distances on Exciton and Trion Binding Energies in van der Waals Heterostructures," *Nano Lett.* **18**, 2725 (2018).

[3] L. Waldecker *et al.*, "Rigid Band Shifts in Two-Dimensional Semiconductors through External Dielectric Screening," *Phys. Rev. Lett.* **123**, 206403 (2019).

[4] A. C. Riis-Jensen, J. Lu, and K. S. Thygesen, "Electrically controlled dielectric band gap engineering in a two-dimensional semiconductor," *Phys. Rev. B* **101**, 121110 (2020).

[5] A. Steinhoff, M. Rösner, F. Jahnke, T. O. Wehling, and C. Gies, "Influence of Excited Carriers on the Optical and Electronic Properties of $MoS_2$," *Nano Lett.*, **14**, 3743 (2014).

[6] Z. Peng, X. Chen, Y. Fan, D. J. Srolovitz, and D. Lei, "Strain engineering of 2D semiconductors and graphene: from strain fields to band-structure tuning and photonic applications," *Light Sci. Appl.* **9**, 190, (2020).

[7] Z. An, M. Zopf, and F. Ding, "Strain-Tuning of 2D Transition Metal Dichalcogenides," in *Nanomembranes*, John Wiley & Sons, Ltd, 413 (2022).

[8] K. F. Mak and J. Shan, "Photonics and optoelectronics of 2D semiconductor transition metal dichalcogenides," *Nat. Photonics* **10**, 216 (2016).

[9] S. J. Liang, B. Cheng, X. Cui, and F. Mao, "Van der Waals heterostructures for high-performance device applications: Challenges and opportunities," *Adv. Mater.* **32**, 1903800 (2020).

[10] C. Gies and A. Steinhoff, "Atomically Thin van der Waals Semiconductors—A Theoretical Perspective," *Laser Photonics Rev.* **15**, 2000482 (2021).

[11] Y. Liu, N. O. Weiss, X. Duan, H.-C. Cheng, Y. Huang, and X. Duan, "Van der Waals heterostructures and devices," *Nat. Rev. Mater.* **1**, 16042 (2016).

[12] J. Choi *et al.*, "Twist Angle-Dependent Interlayer Exciton Lifetimes in van der Waals Heterostructures," *Phys. Rev. Lett.* **126**, 047401 (2021).

[13] J. Hagel, S. Brem, and E. Malic, "Electrical tuning of moiré excitons in $MoSe_2$ bilayers," *2D Mater.* **10**, 014013 (2022).

[14] P. K. Barman *et al.*, "Twist-Dependent Tuning of Excitonic Emissions in Bilayer $WSe_2$," *ACS Omega* **7**, 6412 (2022).

[15] F. Wu, T. Lovorn, and A. H. MacDonald, "Theory of optical absorption by interlayer excitons in transition metal dichalcogenide heterobilayers," *Phys. Rev. B* **97**, 035306 (2018).

[16] K. L. Seyler *et al.*, "Signatures of moiré-trapped valley excitons in $MoSe_2/WSe_2$ heterobilayers," *Nature* **567**, 66 (2019).

[17] S. Brem, C. Linderälv, P. Erhart, and E. Malic, "Tunable Phases of Moiré Excitons in van der Waals Heterostructures," *Nano Lett.* **20**, 8534 (2020).

[18] K. Tran, J. Choi, and A. Singh, "Moiré and beyond in transition metal dichalcogenide twisted bilayers," *2D Mater.* **8**, 022002 (2020).

[19] B. Wu *et al.*, "Evidence for moiré intralayer excitons in twisted $WSe_2/WSe_2$ homobilayer superlattices," *Light Sci. Appl.* **11**, 166 (2022).

[20] H. Yu, G.-B. Liu, J. Tang, X. Xu, and W. Yao, "Moiré excitons: From programmable quantum emitter arrays to spin-orbit–coupled artificial lattices," *Sci. Adv.* **3**, e1701696 (2017).

[21] X. Sun *et al.*, "Enhanced interactions of interlayer excitons in free-standing heterobilayers," *Nature* **610**, 478 (2022).

[22] Z. Sun *et al.*, "Excitonic transport driven by repulsive dipolar interaction in a van der Waals heterostructure," *Nat. Photonics* **16**, 79 (2022).

[23] D. Erkensten, S. Brem, and E. Malic, "Exciton-exciton interaction in transition metal dichalcogenide monolayers and van der Waals heterostructures," *Phys. Rev. B* **103**, 045426 (2021).

[24] W. Li, X. Lu, S. Dubey, L. Devenica, and A. Srivastava, "Dipolar interactions between localized interlayer excitons in van der Waals heterostructures," *Nat. Mater.* **19**, 624 (2020).

[25] K. Tran *et al.*, "Evidence for moiré excitons in van der Waals heterostructures," *Nature* **567**, 71 (2019).

[26] H. Baek *et al.*, "Highly energy-tunable quantum light from moiré-trapped excitons," *Sci. Adv.* **6**, eaba8526 (2020).

[27] G. J. Beirne *et al.*, "Electronic shell structure and carrier dynamics of high aspect ratio InP single quantum dots," *Phys Rev B* **75**, 195302 (2007).

[28] C. Gies, M. Florian, P. Gartner, and F. Jahnke, "The single quantum dot-laser: lasing and strong coupling in the high-excitation regime," *Opt Express* **19**, 14370 (2011).

[29] A. Steinhoff *et al.*, "Combined influence of Coulomb interaction and polarons on the carrier dynamics in InGaAs quantum dots," *Phys. Rev. B* **88**, 205309 (2013).

[30] J. Große, M. von Helversen, A. Koulas-Simos, M. Hermann, and S. Reitzenstein, "Development of site-controlled quantum dot arrays acting as scalable sources of indistinguishable photons," *APL Photonics* **5**, 096107 (2020).

[31] L. Zhang *et al.*, "Highly valley-polarized singlet and triplet interlayer excitons in van der Waals heterostructure," *Phys. Rev. B* **100**, 041402 (2019).

[32] S. Zhao *et al.*, "Excitons in mesoscopically reconstructed moiré heterostructures." *Nat. Nanotechnol.* (2023).

[33] P. Nagler *et al.*, "Interlayer exciton dynamics in a dichalcogenide monolayer heterostructure," *2D Mater.* **4**, 025112 (2017).

[34] Q. Tan, A. Rasmita, Z. Zhang, K. S. Novoselov, and W. Gao, "Signature of Cascade Transitions between Interlayer Excitons in a Moiré Superlattice," *Phys. Rev. Lett.* **129**, 247401 (2022).

[35] A. F. Rigosi, H. M. Hill, Y. Li, A. Chernikov, and T. F. Heinz, "Probing Interlayer Interactions in Transition Metal Dichalcogenide Heterostructures by Optical Spectroscopy: $MoS_2/WS_2$ and $MoSe_2/WSe_2$," *Nano Lett.* **15**, 5033 (2015).

[36] A. Wojs, P. Hawrylak, S. Fafard, and L. Jacak, "Electronic structure and magneto-optics of self-assembled quantum dots," *Phys Rev B* **54**, 5604 (1996).

[37] N. Götting, F. Lohof, and C. Gies, "Moiré-Bose-Hubbard model for interlayer excitons in twisted transition metal

dichalcogenide heterostructures," *Phys. Rev. B* **105**, 165419 (2022).

[38] N. Baer, P. Gartner, and F. Jahnke, "Coulomb effects in semiconductor quantum dots," *Eur. Phys. J. B - Condens. Matter Complex Syst.* **42**, 231 (2004).

[39] A. Barenco and M. A. Dupertuis, "Quantum many-body states of excitons in a small quantum dot," *Phys. Rev. B* **52**, 2766 (1995).

[40] G. Kotliar and D. Vollhardt, "Strongly correlated materials: Insights from dynamical mean-field theory," *Physics Today* **57**, 3, 53 (2004).

[41] C. Lagoin and F. Dubin, "Key role of the moiré potential for the quasicondensation of interlayer excitons in van der Waals heterostructures," *Phys. Rev. B* **103**, L041406 (2021).

[42] A. Julku, "Nonlocal interactions and supersolidity of moiré excitons," *Phys. Rev. B* **106**, 035406, 2022.

[43] B. Laikhtman and R. Rapaport, "Exciton correlations in coupled quantum wells and their luminescence blue shift," *Phys. Rev. B* **80**, 195313 (2009).